\documentclass[aip,jmp,superscriptaddress,preprint]{revtex4-1}
\usepackage{amsmath,amsthm}
\usepackage{amsfonts}
\usepackage[colorlinks=true]{hyperref}
\newtheorem{thm}{Theorem}
\newtheorem{cor}[thm]{Corollary}
\newtheorem{lem}[thm]{Lemma}

\theoremstyle{definition}
\newtheorem{defn}[thm]{Definition}
\theoremstyle{remark}
\newtheorem{rem}[thm]{Remark}
\begin{document}
\title{Multimode Gaussian optimizers for the Wehrl entropy and quantum Gaussian channels}
\author{Giacomo De Palma}
\affiliation{QMATH, Department of Mathematical Sciences, University of Copenhagen, Universitetsparken 5, 2100 Copenhagen, Denmark}
\author{Dario Trevisan}
\affiliation{Universit\`a degli Studi di Pisa, I-56126 Pisa, Italy}
\author{Vittorio Giovannetti}
\affiliation{NEST, Scuola Normale Superiore and Istituto Nanoscienze-CNR, I-56126 Pisa, Italy}

\begin{abstract}
We prove in the multimode scenario a fundamental relation between the Wehrl and the von Neumann entropy, stating that the minimum Wehrl entropy among all the quantum states with a given von Neumann entropy is achieved by thermal Gaussian states.
We also prove that thermal Gaussian input states minimize the output von Neumann entropy of multimode quantum Gaussian attenuators, amplifiers and phase-contravariant channels among all the input states diagonal in some product basis and with a given entropy.
This result constitutes a major step towards the proof of the same property for generic input states, which is still an open conjecture.
This conjecture is necessary to determine the maximum communication rates for the triple trade-off coding and broadcast communication with the Gaussian quantum-limited attenuator.
Finally, we prove that the tensor product of $n$ identical geometric input probability distributions minimizes the output Shannon entropy of the $n$-mode thinning among all the input probability distributions with a given entropy.
\end{abstract}
\maketitle

\section{Introduction}
Electromagnetic waves traveling through cables or free space provide the most promising platform for quantum communication and quantum key distribution \cite{weedbrook2012gaussian}.
Quantum Gaussian channels \cite{holevo2013quantum,holevo2015gaussian} model the propagation of electromagnetic waves in the quantum regime, and can be used to determine the maximum communication and key distribution rates achievable in principle by quantum communication devices.

The determination of the maximum rates for both broadcast communication and the triple trade-off coding with the quantum-limited Gaussian attenuator rely on a constrained minimum output entropy conjecture \cite{guha2007classicalproc,guha2007classical,wilde2012information,wilde2012quantum}.
This conjecture states that Gaussian input states minimize the output von Neumann entropy \cite{holevo2013quantum,wilde2017quantum} of quantum Gaussian channels among all the input states with a given entropy.
This conjecture has been proven only in the one-mode scenario \cite{de2015passive,de2016passive,de2016gaussian,de2016pq,de2016gaussiannew} and is open since 2008.
The only result so far in the multimode scenario is the quantum Entropy Power Inequality \cite{konig2014entropy,konig2016corrections,de2014generalization,de2015multimode}, that provides a lower bound to the output entropy of quantum Gaussian channels.
However, this bound is strictly lower than the output entropy generated by Gaussian input states, and it is not sufficient to prove the conjecture.

The heterodyne measurement \cite{schleich2015quantum} is the most important measurement in quantum optics, and lies at the basis of one of the most promising protocols for quantum key distribution \cite{weedbrook2012gaussian,weedbrook2004quantum}.
The Wehrl entropy \cite{wehrl1979relation,wehrl1978general} of a quantum state is the Shannon differential entropy \cite{cover2006elements} of the probability distribution of the outcome of a heterodyne measurement performed on the state.
A fundamental relation between the Wehrl and the von Neumann entropy has been conjectured \cite{de2017wehrl}.
This relation states that the minimum Wehrl entropy among all the quantum states with a given von Neumann entropy is achieved by Gaussian thermal states, and has been proven only in the one-mode scenario \cite{de2017wehrl}.

In this paper we extend all these results to the multimode scenario.

In \autoref{sec:additivity}, we prove that the constrained minimum output entropy of any classical, quantum-classical and classical-quantum channel is additive (Corollary \ref{cor:cc} and Theorems \ref{thm:qc} and \ref{thm:cq}).
In other words, any lower bound for the output entropy of a classical, quantum-classical or classical-quantum channel implies the corresponding lower bound for the output entropy of the tensor product of $n$ copies of the channel.

In \autoref{sec:gqs}, we present Gaussian quantum systems.

In \autoref{sec:wehrl}, we exploit Theorem \ref{thm:qc} to finally prove in the multimode scenario the conjectured relation between the Wehrl and the von Neumann entropy presented above (Theorem \ref{thm:wehrlmulti}).

In \autoref{sec:gaussian}, we exploit Theorem \ref{thm:cq} to prove that $n$-mode thermal Gaussian input states minimize the output von Neumann entropy of the tensor product of $n$ identical one-mode phase-covariant or -contravariant quantum Gaussian channels among all the input states diagonal in some product basis and with a given entropy (Theorem \ref{thm:epnimulti}).
In particular, this result holds for any input state diagonal in the joint Fock basis.

In \autoref{sec:thinning}, we exploit Corollary \ref{cor:cc} to prove that the tensor product of $n$ identical geometric distributions minimizes the output Shannon entropy of the $n$-mode thinning among all the input probability distributions with a given entropy (Theorem \ref{thmthinmulti}).
The thinning \cite{renyi1956characterization,de2015passive} is the map on classical probability measures on $\mathbb{N}$ that is the discrete equivalent of the rescaling of a positive real random variable.
It has been involved in discrete versions of the central limit theorem \cite{harremoes2007thinning,yu2009monotonic,harremoes2010thinning}
and of the Entropy Power Inequality \cite{yu2009concavity,johnson2010monotonicity}.

In \autoref{sec:concl} we conclude.

Appendix \ref{app:lem} contains the proof of the auxiliary Lemmas.

\section{Additivity of constrained minimum output entropy}\label{sec:additivity}
\begin{thm}[CMOE additivity: quantum-classical channels]\label{thm:qc}
Let $A$ be a quantum system with Hilbert space $\mathcal{H}_A$, and let $X$ be a countable set.
Let $\Phi$ be a quantum-classical channel that sends quantum states on $\mathcal{H}_A$ to probability measures on $X$, representing a measurement on $A$ with outcomes in $X$.
Then, the constrained minimum output entropy of $\Phi$ is additive.
In other words, let us suppose that for any quantum state $\hat{\rho}_A$ on $\mathcal{H}_A$
\begin{equation}\label{eq:hyp}
H\left(\Phi\left(\hat{\rho}_A\right)\right) \ge f\left(S\left(\hat{\rho}_A\right)\right)\;,
\end{equation}
where $H$ is the Shannon entropy, $S$ is the von Neumann entropy and $f:[0,\infty)\to[0,\infty)$ is increasing and convex.
Then, for any $n\in\mathbb{N}$ and any quantum state $\hat{\rho}_{A_1\ldots A_n}$ on $\mathcal{H}_A^{\otimes n}$
\begin{equation}
H\left(\Phi^{\otimes n}\left(\hat{\rho}_{A_1\ldots A_n}\right)\right) \ge n\;f\left(\left.S\left(\hat{\rho}_{A_1\ldots A_n}\right)\right/n\right)\;.
\end{equation}
\begin{proof}
We will prove the claim by induction on $n$.
The claim is true for $n=1$.
From the inductive hypothesis, we can assume the claim for a given $n$.
It is then sufficient to prove the claim for $n+1$.
Given a quantum state $\hat{\rho}_{A_1\ldots A_{n+1}}$ on $\mathcal{H}_A^{\otimes n+1}$, $\Phi^{\otimes n+1}\left(\hat{\rho}_{A_1\ldots A_{n+1}}\right)$ is the probability distribution of the outcomes $X_1\ldots X_{n+1}$ of the measurements on $A_1\ldots A_{n+1}$.
The idea of the proof is applying the chain rule for the entropy and conditioning on the values of $X_1\ldots X_n$:
\begin{align}\label{eq:chain}
H(X_1\ldots X_{n+1}) &= H(X_1\ldots X_n) + H(X_{n+1}|X_1\ldots X_n)\nonumber\\
&= H(X_1\ldots X_n) + \int_{X_1\ldots X_n} H(X_{n+1}|X_1=x_1\ldots X_n=x_n)\mathrm{d}p_{X_1\ldots X_n}(x_1\ldots x_n)\,,
\end{align}
where all the entropies are computed on $\Phi^{\otimes n+1}\left(\hat{\rho}_{A_1\ldots A_{n+1}}\right)$, and $p_{X_1\ldots X_n}$ is the probability measure on $X_1\ldots X_n$ induced by $\Phi^{\otimes n+1}\left(\hat{\rho}_{A_1\ldots A_{n+1}}\right)$.
$H(X_{n+1}|X_1=x_1\ldots X_n=x_n)$ is the entropy of $\Phi(\hat{\rho}_{A_{n+1}|X_1=x_1\ldots X_n=x_n})$, where $\hat{\rho}_{A_{n+1}|X_1=x_1\ldots X_n=x_n}$ is the reduced state of $A_{n+1}$ conditioned on the outcomes $x_1\ldots x_n$ for the measurements on $A_1\ldots A_n$.
We have from the hypothesis \eqref{eq:hyp}
\begin{equation}\label{eq:SX}
S(\Phi(\hat{\rho}_{A_{n+1}|X_1=x_1\ldots X_n=x_n})) \ge f(S(\hat{\rho}_{A_{n+1}|X_1=x_1\ldots X_n=x_n}))\;.
\end{equation}
Applying \eqref{eq:SX} and the inductive hypothesis to \eqref{eq:chain} we get
\begin{align}\label{eq:ind}
H(X_1\ldots X_{n+1}) &\ge n\,f(S(A_1\ldots A_n)/n)\nonumber\\
& \phantom{\ge} + \int_{X_1\ldots X_n} f(S(A_{n+1}|X_1=x_1\ldots X_n=x_n))\,\mathrm{d}p_{X_1\ldots X_n}(x_1\ldots x_n)\;.
\end{align}
Since $f$ is convex we have
\begin{align}\label{eq:conv}
&\int_{X_1\ldots X_n} f(S(A_{n+1}|X_1=x_1\ldots X_n=x_n))\,\mathrm{d}p_{X_1\ldots X_n}(x_1\ldots x_n)\nonumber\\
&\ge f\left(\int_{X_1\ldots X_n} S(A_{n+1}|X_1=x_1\ldots X_n=x_n)\,\mathrm{d}p_{X_1\ldots X_n}(x_1\ldots x_n)\right) = f(S(A_{n+1}|X_1\ldots X_n))\;.
\end{align}
From the data-processing inequality for the conditional entropy we have
\begin{equation}
S(A_{n+1}|X_1\ldots X_n) \ge S(A_{n+1}|A_1\ldots A_n)\;,
\end{equation}
and since $f$ is increasing we also have
\begin{equation}\label{eq:incr}
f(S(A_{n+1}|X_1\ldots X_n)) \ge f(S(A_{n+1}|A_1\ldots A_n))\;.
\end{equation}
Finally, putting together \eqref{eq:ind}, \eqref{eq:conv} and \eqref{eq:incr} we get
\begin{align}
H(X_1\ldots X_{n+1}) &\ge n\,f(S(A_1\ldots A_n)/n) + f(S(A_{n+1}|A_1\ldots A_n))\nonumber\\
&\ge \left(n+1\right)f\left(\frac{S(A_1\ldots A_n)+S(A_{n+1}|A_1\ldots A_n)}{n+1}\right)\nonumber\\
&= \left(n+1\right)f\left(\left.S(A_1\ldots A_{n+1})\right/\left(n+1\right)\right)\;,
\end{align}
where we have used the convexity of $f$ again.
\end{proof}
\end{thm}
\begin{thm}[CMOE additivity: classical-quantum channels]\label{thm:cq}
Let $X$ be a countable set, and let $A$ be a quantum system with Hilbert space $\mathcal{H}_A$.
Let $\Phi$ be a classical-quantum channel that sends probability measures on $X$ to quantum states on $\mathcal{H}_A$.
Then, the constrained minimum output entropy of $\Phi$ is additive.
In other words, let us suppose that for any probability measure $p_X$ on $X$
\begin{equation}\label{eq:cq1}
S\left(\Phi\left(p_X\right)\right) \ge f\left(H\left(p_X\right)\right)\;,
\end{equation}
where $f:[0,\infty)\to[0,\infty)$ is increasing and convex.
Then, for any $n\in\mathbb{N}$ and any probability measure $p_{X_1\ldots X_n}$ on $n$ copies of $X$,
\begin{equation}
S\left(\Phi^{\otimes n}\left(p_{X_1\ldots X_n}\right)\right) \ge n\;f\left(\left.H\left(p_{X_1\ldots X_n}\right)\right/n\right)\;.
\end{equation}
\begin{proof}
We will prove the claim by induction on $n$.
The claim is true for $n=1$.
From the inductive hypothesis, we can assume the claim for a given $n$.
It is then sufficient to prove the claim for $n+1$.
Let us then consider a probability measure $p_{X_1\ldots X_{n+1}}$ on $n+1$ copies of $X$.
As in the proof of Theorem \ref{thm:qc}, we will use the chain rule for the entropy and condition on the values of $X_1\ldots X_n$:
\begin{align}
&S(A_1\ldots A_{n+1}) = S(A_1\ldots A_n) + S(A_{n+1}|A_1\ldots A_n)\nonumber\\
&\overset{(a)}{\ge} S(A_1\ldots A_n) + S(A_{n+1}|X_1\ldots X_n)\nonumber\\
&= S(A_1\ldots A_n) + \int_{X_1\ldots X_n}S(A_{n+1}|X_1=x_1\ldots X_n=x_n)\,\mathrm{d}p_{X_1\ldots X_n}(x_1\ldots x_n)\nonumber\\
&\overset{(b)}{\ge} n\,f(H(X_1\ldots X_n)/n) + \int_{X_1\ldots X_n}f(H(X_{n+1}|X_1=x_1\ldots X_n=x_n))\,\mathrm{d}p_{X_1\ldots X_n}(x_1\ldots x_n)\nonumber\\
&\overset{(c)}{\ge} n\,f(H(X_1\ldots X_n)/n) + f\left(\int_{X_1\ldots X_n}H(X_{n+1}|X_1=x_1\ldots X_n=x_n)\right)\,\mathrm{d}p_{X_1\ldots X_n}(x_1\ldots x_n)\nonumber\\
&= n\,f(H(X_1\ldots X_n)/n) + f(H(X_{n+1}|X_1\ldots X_n))\nonumber\\
&\overset{(d)}{\ge} \left(n+1\right)f\left(\frac{H(X_1\ldots X_n) + H(X_{n+1}|X_1\ldots X_n)}{n+1}\right) = \left(n+1\right)f\left(\frac{H(X_1\ldots X_{n+1})}{n+1}\right)\;,
\end{align}
where $p_{X_1\ldots X_n}$ is the marginal of $p_{X_1\ldots X_{n+1}}$, $(a)$ follows from the data-processing inequality for the conditional entropy, $(b)$ follows from the inductive hypothesis and from the claim for $n=1$, and $(c)$ and $(d)$ follow from the convexity of $f$.
\end{proof}
\end{thm}
\begin{rem}\label{rem:diff}
Theorems \ref{thm:qc} and \ref{thm:cq} hold also with $X=\mathbb{R}^m$ and the Shannon entropy replaced by the Shannon differential entropy
\begin{equation}
H(X):=-\int_{\mathbb{R}^m}\ln p_X(x)\;\mathrm{d}p_X(x)\;,
\end{equation}
where $p_X(x)$ is the probability density function of $X$.
\end{rem}

With an identical proof we get
\begin{cor}\label{cor:cq}
Let $X_1\ldots X_n$ be countable sets, and let $A_1\ldots A_n$ be quantum systems.
Let $\Phi_1\ldots\Phi_n$ be classical-quantum channels such that each $\Phi_i$ sends probability measures on $X_i$ to quantum states on $A_i$.
Let us suppose that for any $i=1,\ldots,\,n$ and any probability measure $p_{X_i}$ on $X_i$
\begin{equation}
S\left(\Phi_i\left(p_{X_i}\right)\right) \ge f\left(H\left(p_{X_i}\right)\right)\;,
\end{equation}
where $f$ is increasing and convex and does not depend on $i$.
Then, for any probability measure $p_{X_1\ldots X_n}$ on $X_1\ldots X_n$,
\begin{equation}
S\left(\left(\Phi_1\otimes\ldots\otimes\Phi_n\right)\left(p_{X_1\ldots X_n}\right)\right) \ge n\;f\left(\left.H\left(p_{X_1\ldots X_n}\right)\right/n\right)\;.
\end{equation}
\end{cor}
Theorems \ref{thm:qc} and \ref{thm:cq} hold in particular when both the input and the output are classical:
\begin{cor}[CMOE additivity: classical channels]\label{cor:cc}
Let $X$ and $Y$ be countable sets.
Let $\Phi$ be a classical channel that sends probability measures on $X$ to probability measures on $Y$.
Then, the constrained minimum output entropy of $\Phi$ is additive.
In other words, let us suppose that for any probability measure $p_X$ on $X$
\begin{equation}
H\left(\Phi\left(p_X\right)\right) \ge f\left(H\left(p_X\right)\right)\;,
\end{equation}
where $f:[0,\infty)\to[0,\infty)$ is increasing and convex.
Then, for any $n\in\mathbb{N}$ and any probability measure $p_{X_1\ldots X_n}$ on $n$ copies of $X$,
\begin{equation}
H\left(\Phi^{\otimes n}\left(p_{X_1\ldots X_n}\right)\right) \ge n\;f\left(\left.H\left(p_{X_1\ldots X_n}\right)\right/n\right)\;.
\end{equation}
\end{cor}

\section{Gaussian quantum systems}\label{sec:gqs}
We consider the Hilbert space of $n$ harmonic oscillators, or $n$ modes of the electromagnetic radiation.
The ladder operators $\hat{a}_1\ldots\hat{a}_n$ satisfy the bosonic canonical commutation relations
\begin{equation}
\left[\hat{a}_i,\;\hat{a}^\dag_j\right]=\delta_{ij}\;\hat{\mathbb{I}}\;,\qquad i,\,j=1,\ldots,\,n\;,
\end{equation}
and the Hamiltonian
\begin{equation}\label{eq:defH}
\hat{H}:=\sum_{i=1}^n\hat{a}_i^\dag\hat{a}_i
\end{equation}
counts the total number of excitations, or photons.
The operators $\hat{a}^\dag_1\hat{a}_1\ldots \hat{a}^\dag_n\hat{a}_n$ can be simultaneously diagonalized, and the resulting eigenbasis is the joint Fock basis
\begin{equation}
\left\{|k_1\ldots k_n\rangle\right\}_{k_1\ldots k_n\in\mathbb{N}}\;.
\end{equation}

Thermal Gaussian states are the Gibbs state of the Hamiltonian \eqref{eq:defH}.
For one mode ($n=1$), the density matrix of the thermal Gaussian state with average energy $E\ge0$ is
\begin{equation}\label{eq:omegaz}
\hat{\omega}_E=\sum_{k=0}^\infty \frac{1}{E+1}\left(\frac{E}{E+1}\right)^k|k\rangle\langle k|\;,
\end{equation}
where $\{|k\rangle\}_{k\in\mathbb{N}}$ is the Fock basis.
Its von Neumann entropy is
\begin{equation}\label{eq:Sz}
S\left(\hat{\omega}_E\right)= \left(E+1\right)\ln\left(E+1\right)-E\ln E:=g(E)\;.
\end{equation}
The $n$-mode thermal Gaussian state with average energy per mode $E$ is $\hat{\omega}_E^{\otimes n}$.

\section{Gaussian optimizers for the Wehrl entropy}\label{sec:wehrl}
The classical phase space associated to an $n$-mode Gaussian quantum system is $\mathbb{C}^n$, and for any $\mathbf{z}\in\mathbb{C}^n$ we define the coherent state
\begin{equation}
|\mathbf{z}\rangle = e^{-\frac{|\mathbf{z}|^2}{2}}\sum_{k_1,\ldots,\,k_M\in\mathbb{N}}\frac{z_1^{k_1}\ldots z_M^{k_M}}{\sqrt{k_1!\ldots k_M!}}\;|k_1\ldots k_M\rangle\;.
\end{equation}
Coherent states are not orthogonal:
\begin{equation}\label{eq:cohov}
\langle\mathbf{z}|\mathbf{w}\rangle = e^{\mathbf{z}^\dag\mathbf{w}-\frac{|\mathbf{z}|^2+|\mathbf{w}|^2}{2}}\qquad\forall\;\mathbf{z},\,\mathbf{w}\in\mathbb{C}^n\;,
\end{equation}
but they are complete and satisfy the resolution of the identity \cite{holevo2015gaussian}
\begin{equation}\label{eq:complz}
\int_{\mathbb{C}^n}|\mathbf{z}\rangle\langle \mathbf{z}|\;\frac{\mathrm{d}^{2n}z}{\pi^n} = \hat{\mathbb{I}}\;,
\end{equation}
where the integral converges in the weak topology.
\begin{defn}[heterodyne measurement \cite{schleich2015quantum}]
The heterodyne measurement is the POVM associated with the resolution of the identity \eqref{eq:complz}, that sends the quantum state $\hat{\rho}$ to the probability measure on $\mathbb{C}^n$ with density
\begin{equation}
\mathrm{d}p(\mathbf{z}) = \langle\mathbf{z}|\hat{\rho}|\mathbf{z}\rangle\;\frac{\mathrm{d}^{2n}z}{\pi^n}\;.
\end{equation}
\end{defn}
\begin{defn}[Wehrl entropy]
The Wehrl entropy of a quantum state $\hat{\rho}$ is the Shannon differential entropy of the probability distribution of the outcome of an heterodyne measurement performed on $\hat{\rho}$:
\begin{equation}
W\left(\hat{\rho}\right) := -\int_{\mathbb{C}^n} \langle\mathbf{z}|\hat{\rho}|\mathbf{z}\rangle\;\ln \langle\mathbf{z}|\hat{\rho}|\mathbf{z}\rangle\;\frac{\mathrm{d}^{2n}z}{\pi^n}\;.
\end{equation}
\end{defn}
The fundamental relation between the Wehrl and the von Neumann entropy has been proven in Ref. \cite{de2017wehrl} in the one-mode scenario:
\begin{thm}[one-mode Wehrl entropy has Gaussian optimizers \cite{de2017wehrl}]\label{thm:wehrlone}
The minimum Wehrl entropy among all the one-mode quantum states with a given von Neumann entropy is achieved by thermal Gaussian states, i.e. for any one-mode quantum state $\hat{\rho}$
\begin{equation}\label{eq:entropy}
W\left(\hat{\rho}\right) \ge \ln\left(g^{-1}\left(S\left(\hat{\rho}\right)\right)+1\right)+1\;,
\end{equation}
with $g$ defined in \eqref{eq:Sz}.
\end{thm}
Thanks to Theorem \ref{thm:qc} we can finally extend this result to the multimode scenario:
\begin{thm}[multimode Wehrl entropy has Gaussian optimizers]\label{thm:wehrlmulti}
For any $n\in\mathbb{N}$, the minimum Wehrl entropy among all the $n$-mode quantum states with a given von Neumann entropy is achieved by thermal Gaussian states, i.e. for any $n$-mode quantum state $\hat{\rho}$
\begin{equation}\label{eq:entropymulti}
W\left(\hat{\rho}\right) \ge n\left(\ln\left(g^{-1}\left(\left.S\left(\hat{\rho}\right)\right/n\right)+1\right)+1\right)\;.
\end{equation}
\begin{proof}
The claim follows from Theorem \ref{thm:qc} and Remark \ref{rem:diff} applied to the heterodyne measurement, together with Theorem \ref{thm:wehrlone} and Lemma \ref{lem:fw} of Appendix \ref{app:lem}.
\end{proof}
\end{thm}

\section{Gaussian optimizers for quantum Gaussian channels}\label{sec:gaussian}
The one-mode Gaussian quantum attenuator $\mathcal{E}_{\lambda,E}$ of transmissivity $0\le\lambda\le 1$ and thermal energy $E\ge 0$ mixes the input state $\hat{\rho}$ with the thermal Gaussian state $\hat{\omega}_E$ of an environmental quantum system $B$ through a beamsplitter of transmissivity $\lambda$
(\cite{holevo2007one}, case (C) with $k=\sqrt{\lambda}$ and $N_0=E$):
\begin{equation}
\mathcal{E}_{\lambda,E}\left(\hat{\rho}\right)=\mathrm{Tr}_B\left[\hat{U}_\lambda\left(\hat{\rho}\otimes\hat{\omega}_E\right)\hat{U}_\lambda^\dag\right]\;.
\end{equation}
Here $\mbox{Tr}_B[ \cdots]$ is the partial trace over the environment $B$,
\begin{equation}
\hat{U}_\lambda=\exp\left(\left(\hat{a}^\dag\hat{b}-\hat{a}\,\hat{b}^\dag\right)\arccos\sqrt{\lambda}\right)
\end{equation}
is the unitary operator implementing the beamsplitter, and $\hat{b}$ is the ladder operator of $B$ (\cite{ferraro2005gaussian}, Section 1.4.2).
For $E=0$ the state of the environment is the vacuum and the attenuator is quantum-limited.
We put $\mathcal{E}_{\lambda,0}=\mathcal{E}_\lambda$ for simplicity.
The action of the quantum Gaussian attenuator on thermal Gaussian states is~\cite{holevo2013quantum}
\begin{equation} \label{TRANSFGAUS}
\mathcal{E}_{\lambda,E}\left(\hat{\omega}_{E'}\right)=\hat{\omega}_{\lambda E'+(1-\lambda)E}\;.
\end{equation}
The $n$-mode quantum Gaussian attenuator of transmissivity $0\le\lambda\le 1$ and thermal energy per mode $E\ge 0$ is $\mathcal{E}_{\lambda,E}^{\otimes n}$, the tensor product of $n$ identical one-mode quantum Gaussian attenuators.

The quantum Gaussian amplifier $\mathcal{A}_{\kappa,E}$ of amplification parameter $\kappa\ge1$ and thermal energy $E\ge 0$ performs a two-mode squeezing on the input state $\hat{\rho}$ and the thermal Gaussian state $\hat{\omega}_E$ of  $B$ (\cite{holevo2007one}, case (C) with $k=\sqrt{\kappa}$ and $N_0=E$):
\begin{equation}
\mathcal{A}_{\kappa,E}\left(\hat{\rho}\right)=\mathrm{Tr}_B\left[\hat{U}_\kappa\left(\hat{\rho}\otimes \hat{\omega}_E\right)\hat{U}_\kappa^\dag\right]\;,
\end{equation}
where
\begin{equation}\label{eq:defUk}
\hat{U}_\kappa=\exp\left(\left(\hat{a}^\dag\hat{b}^\dag-\hat{a}\,\hat{b}\right)\mathrm{arccosh}\sqrt{\kappa}\right)
\end{equation}
is the squeezing  unitary operator.
Again for $E=0$ the amplifier is quantum-limited and we put $\mathcal{A}_{\kappa,0}=\mathcal{A}_\kappa$ for simplicity.
The action of the quantum Gaussian amplifier on thermal Gaussian states is
\begin{equation}\label{eq:AE'}
\mathcal{A}_{\kappa,E}\left(\hat{\omega}_{E'}\right)=\hat{\omega}_{\kappa E'+(\kappa-1)(E+1)}\;.
\end{equation}
The $n$-mode quantum Gaussian amplifier of amplification parameter $\kappa\ge1$ and thermal energy per mode $E\ge 0$ is $\mathcal{A}_{\kappa,E}^{\otimes n}$, the tensor product of $n$ identical one-mode quantum Gaussian amplifiers.

The $n$-mode quantum Gaussian additive-noise channel $\mathcal{N}_E^{\otimes n}$ adds $E\ge0$ to the energy of each mode of the input state (\cite{holevo2007one}, case ($\text{B}_2$) with $N_c=E$).
For $n=1$, its action on thermal Gaussian states is
\begin{equation}
\mathcal{N}_E\left(\hat{\omega}_{E'}\right) = \hat{\omega}_{E'+E}\;.
\end{equation}

The one-mode phase-contravariant quantum Gaussian channel $\tilde{\mathcal{A}}_{\kappa,E}$ is the weak complementary of the one-mode quantum Gaussian amplifier $\mathcal{A}_{\kappa,E}$ (\cite{holevo2007one}, case (D) with $k=\sqrt{\kappa-1}$ and $N_0=E$):
\begin{equation}
\tilde{\mathcal{A}}_{\kappa,E}\left(\hat{\rho}\right)=\mathrm{Tr}_A\left[\hat{U}_\kappa\left(\hat{\rho}\otimes \hat{\omega}_E\right)\hat{U}_\kappa^\dag\right]\;,
\end{equation}
where $\hat{U}_\kappa$ is the two-mode squeezing unitary defined in \eqref{eq:defUk} and where now the partial trace is performed over the system $A$.
The action of $\tilde{\mathcal{A}}_{\kappa,E}$ on thermal Gaussian states is \cite{holevo2013quantum}
\begin{equation}
\tilde{\mathcal{A}}_{\kappa,E}\left(\hat{\omega}_{E'}\right)=\hat{\omega}_{(\kappa-1)(E'+1)+\kappa E}\;.
\end{equation}
The $n$-mode phase-contravariant quantum Gaussian channel $\tilde{\mathcal{A}}_{\kappa,E}^{\otimes n}$ is the tensor product of $n$ identical one-mode phase-contravariant quantum Gaussian channels.

A fundamental property of the output entropy of one-mode quantum Gaussian channels has been proven in Ref. \cite{de2016gaussiannew}:
\begin{thm}[one-mode CMOE conjecture \cite{de2016gaussiannew}]\label{thm:epnione}
Gaussian thermal input states minimize the output von Neumann entropy of any one-mode phase-covariant and -contravariant quantum Gaussian channel among all the input states with a given entropy, i.e.\ for any $0\le\lambda\le1$, $\kappa\ge1$, $E\ge0$ and any quantum state $\hat{\rho}$
\begin{subequations}
\begin{align}
S\left(\mathcal{E}_{\lambda,E}\left(\hat{\rho}\right)\right) &\ge g\left(\lambda\;g^{-1}\left(S\left(\hat{\rho}\right)\right)+\left(1-\lambda\right)E\right)\;,\\
S\left(\mathcal{N}_E\left(\hat{\rho}\right)\right) &\ge g\left(g^{-1}\left(S\left(\hat{\rho}\right)\right)+E\right)\;,\\
S\left(\mathcal{A}_{\kappa,E}\left(\hat{\rho}\right)\right) &\ge g\left(\kappa\;g^{-1}\left(S\left(\hat{\rho}\right)\right)+\left(\kappa-1\right)\left(E+1\right)\right)\\
S\left(\tilde{\mathcal{A}}_{\kappa,E}\left(\hat{\rho}\right)\right) &\ge g\left(\left(\kappa-1\right)\left(g^{-1}\left(S\left(\hat{\rho}\right)\right)+1\right)+\kappa\;E\right)\;.
\end{align}
\end{subequations}
\end{thm}
Thanks to Theorem \ref{thm:cq} we can extend this result to the multimode scenario:
\begin{thm}[multimode CMOE conjecture]\label{thm:epnimulti}
For any $n\in\mathbb{N}$, Gaussian thermal input states minimize the output von Neumann entropy of the $n$-mode quantum Gaussian attenuator, amplifier, additive-noise channel and phase-contravariant channel among all the input states diagonal in some product basis and with a given entropy.
In other words, given $n$ orthonormal bases $\left\{\left|\psi^{(1)}_x\right\rangle\right\}_{x\in\mathbb{N}},\;\ldots,\;\left\{\left|\psi^{(n)}_x\right\rangle\right\}_{x\in\mathbb{N}}$ of the Hilbert space of one harmonic oscillator, for any quantum state $\hat{\rho}$ of the form
\begin{equation}\label{eq:rhop}
\hat{\rho} = \sum_{x_1\ldots x_n\in\mathbb{N}} p_{X_1\ldots X_n}(x_1\ldots x_n)\;\left|\psi^{(1)}_{x_1}\ldots\psi^{(n)}_{x_n}\right\rangle\left\langle\psi^{(1)}_{x_1}\ldots\psi^{(n)}_{x_n}\right|\;,
\end{equation}
where $p_{X_1\ldots X_n}$ is a probability measure on $\mathbb{N}^n$, and for any $0\le\lambda\le1$, $\kappa\ge1$ and $E\ge0$
\begin{subequations}\label{eq:epni}
\begin{align}\label{eq:epniatt}
S\left(\mathcal{E}_{\lambda,E}^{\otimes n}\left(\hat{\rho}\right)\right) &\ge n\;g\left(\lambda\;g^{-1}\left(\left.S\left(\hat{\rho}\right)\right/n\right)+\left(1-\lambda\right)E\right)\;,\\
S\left(\mathcal{N}_E^{\otimes n}\left(\hat{\rho}\right)\right) &\ge n\;g\left(g^{-1}\left(\left.S\left(\hat{\rho}\right)\right/n\right)+E\right)\;,\\
S\left(\mathcal{A}_{\kappa,E}^{\otimes n}\left(\hat{\rho}\right)\right) &\ge n\;g\left(\kappa\;g^{-1}\left(\left.S\left(\hat{\rho}\right)\right/n\right)+\left(\kappa-1\right)\left(E+1\right)\right)\;,\\
S\left(\tilde{\mathcal{A}}_{\kappa,E}^{\otimes n}\left(\hat{\rho}\right)\right) &\ge n\;g\left(\left(\kappa-1\right)\left(g^{-1}\left(\left.S\left(\hat{\rho}\right)\right/n\right)+1\right)+\kappa\;E\right)\;.
\end{align}
\end{subequations}
In particular, Eqs. \eqref{eq:epni} hold for any quantum state $\hat{\rho}$ diagonal in the joint Fock basis.
\begin{proof}
Let us consider the quantum Gaussian attenuator.
The claim \eqref{eq:epniatt} follows from Corollary \ref{cor:cq} applied to the classical-quantum channels
\begin{equation}
\Phi_i(p_X) = \mathcal{E}_{\lambda,E}\left(\sum_{x\in\mathbb{N}}p_X(x)\;\left|\psi^{(i)}_x\right\rangle\left\langle\psi^{(i)}_x\right|\right)\;,\qquad i=1,\,\ldots,\,n\;,
\end{equation}
together with Theorem \ref{thm:epnione} and Lemma \ref{lem:f} of Appendix \ref{app:lem}, noticing that the von Neumann entropy of the state \eqref{eq:rhop} coincides with the Shannon entropy of $p_{X_1\ldots X_n}$.
The proof for the other channels is identical.
\end{proof}
\end{thm}

\section{Thinning}\label{sec:thinning}
The thinning \cite{renyi1956characterization} is the map acting on probability measures on $\mathbb{N}$ that is the discrete analogue of the rescaling on $\mathbb{R}^+$.
\begin{defn}[thinning]
Let $N$ be a random variable with values in $\mathbb{N}$.
The thinning with parameter $0\leq\lambda\leq1$ is defined as
\begin{equation}
T_\lambda(N)=\sum_{i=1}^N B_i\;,
\end{equation}
where the $B_i$ are independent Bernoulli variables with parameter $\lambda$, i.e. each $B_i$ is one with probability $\lambda$, and zero with probability $1-\lambda$.
\end{defn}
The thinning can be understood as follows.
Let us consider a barrier that transmits a particle with probability $\lambda$ and reflects it with probability $1-\lambda$.
Let $N$ be the random variable associated to the number of incoming particles, and $p_N$ its probability distribution.
Then, $T_\lambda(p_N)$ is the probability distribution of the number of transmitted particles.

A fundamental property of the output Shannon entropy of the thinning is \cite{de2016gaussian}
\begin{thm}[CMOE of thinning]\label{thmthin}
Geometric input probability distributions minimize the output Shannon entropy of the thinning among all the input probability distributions with a given entropy, i.e. for any probability distribution $p$ on $\mathbb{N}$ and any $0\leq\lambda\leq 1$
\begin{equation}
H\left(T_\lambda(p)\right)\geq g\left(\lambda\;g^{-1}\left(H(p)\right)\right)\;.
\end{equation}
\end{thm}

The $n$-mode thinning $T_\lambda^{\otimes n}$ is the tensor product of $n$ copies of the thinning, i.e. the map acting on probability distributions on $\mathbb{N}^n$ that applies the thinning independently on each component.
Thanks to Corollary \ref{cor:cc}, we can extend Theorem \ref{thmthin} to the multimode scenario:
\begin{thm}[CMOE of multimode thinning]\label{thmthinmulti}
For any $n\in\mathbb{N}$, the tensor product of $n$ identical geometric input probability distributions minimizes the output Shannon entropy of the $n$-mode thinning among all the input probability distributions with a given entropy, i.e. for any probability distribution $p$ on $\mathbb{N}^n$ and any $0\leq\lambda\leq 1$
\begin{equation}
H\left(T_\lambda^{\otimes n}(p)\right)\geq n\;g\left(\lambda\;g^{-1}\left(\left.H(p)\right/n\right)\right)\;.
\end{equation}
\begin{proof}
The claim follows from Corollary \ref{cor:cc}, Theorem \ref{thmthin} and Lemma \ref{lem:f} of Appendix \ref{app:lem}.
\end{proof}
\end{thm}

\section{Conclusions}\label{sec:concl}
With Theorem \ref{thm:wehrlmulti}, we have finally proven the fundamental relation between the Wehrl and the von Neumann entropies in the multimode scenario.

With Theorem \ref{thm:epnimulti}, we have proven the constrained minimum output entropy conjecture for multimode quantum Gaussian channels for input states diagonal in some product basis.
This result provides a further strong evidence to the validity of the conjecture for generic input states, that will be investigated in future work.

\section*{Acknowledgements}
GdP acknowledges financial support from the European Research Council (ERC Grant Agreement no 337603), the Danish Council for Independent Research (Sapere Aude) and VILLUM FONDEN via the QMATH Centre of Excellence (Grant No. 10059).

\appendix

\section{}\label{app:lem}
\begin{lem}\label{lem:f}
For any $b\ge0$ and any $0\le a\le b+1$ the function
\begin{equation}
f(x):=g\left(a\,g^{-1}(x) + b\right)\;,\qquad x\ge0
\end{equation}
is increasing and convex\;.
\begin{proof}
Since $g$ is increasing, also $g^{-1}$ is increasing, hence $f$ is increasing.
We will prove that $f''(x)\ge0$ for any $x>0$.
We have for any $y>0$
\begin{equation}\label{eq:f''}
f''(g(y)) = \frac{a\;g''(a\,y+b)\;g''(y)}{{g'(y)}^3}\left(a\,h(y) - h(a\,y+b)\right)\;,
\end{equation}
where
\begin{equation}
h(t):=\frac{g'(t)}{g''(t)} = t\left(t+1\right)\ln\frac{t}{t+1}\;,\qquad t>0\;.
\end{equation}
Since $g$ is increasing and concave, the first factor in the right-hand side of \eqref{eq:f''} is positive.
It is then sufficient to prove that
\begin{equation}\label{eq:claimh}
h(a\,y+b) \overset{?}{\le} a\,h(y)\;.
\end{equation}
We consider the change of variable
\begin{equation}
t = \frac{1}{e^\theta-1}\;,\qquad \theta = \ln\frac{t+1}{t}>0\;.
\end{equation}
We have
\begin{equation}
h'(t) = 1-\frac{\theta}{\tanh\frac{\theta}{2}} \le -1 < 0\;,
\end{equation}
hence $h$ is decreasing.
\begin{itemize}
\item If $0\le a\le1$, we consider the function
\begin{equation}
\phi(t) := \frac{h(t)}{t} = \left(t+1\right)\ln\frac{t}{t+1}\;.
\end{equation}
We have
\begin{equation}
\phi'(t) = e^\theta - \theta - 1 \ge 0\;,
\end{equation}
hence $\phi$ is increasing, and
\begin{equation}
\phi(a\,y)\le\phi(y)\;.
\end{equation}
Recalling that $h$ is decreasing, we have
\begin{equation}
h(a\,y+b) \le h(a\,y) \le a\,h(y)\;,
\end{equation}
and the claim \eqref{eq:claimh} follows.
\item If $a>1$, we define the function
\begin{equation}
\phi(t) := \frac{h(t)}{t+1} = t\ln\frac{t}{t+1}\;.
\end{equation}
We have
\begin{equation}
\phi'(t) = 1 - \theta - e^{-\theta} \le 0\;,
\end{equation}
hence $\phi$ is decreasing, and
\begin{equation}
\phi(a\,y+a-1) \le \phi(y)\;.
\end{equation}
Recalling that $h$ is decreasing and $b\ge a-1$ for hypothesis, we have
\begin{equation}
h(a\,y+b) \le h(a\,y+a-1) \le a\,h(y)\;,
\end{equation}
and the claim \eqref{eq:claimh} follows.
\end{itemize}
\end{proof}
\end{lem}
\begin{lem}\label{lem:fw}
The function
\begin{equation}
f(x):= \ln\left(g^{-1}(x) + 1\right) + 1\;,\qquad x\ge0
\end{equation}
is increasing and convex.
\begin{proof}
The claim follows from Lemma \ref{lem:f}, since for any $x\ge0$
\begin{equation}
\ln\left(g^{-1}(x) + 1\right) + 1 = \lim_{\kappa\to\infty}\left(g\left(\kappa\,g^{-1}(x)+\kappa-1\right)-\ln\kappa\right)\;.
\end{equation}
\end{proof}
\end{lem}

\bibliography{biblio}
\bibliographystyle{plain}

\end{document}